


\font\twelverm=cmr12
\font\twelvei=cmmi12
\font\twelvesy=cmsy10 scaled 1200
\font\twelvebf=cmbx12
\font\twelvett=cmtt12
\font\twelveit=cmti12
\font\twelvesl=cmsl12
\font\twelvesc=cmcsc10 scaled 1200
\font\twelvess=cmss12
\font\twelvessi=cmssi12
\font\twelvebmit=cmmib10 scaled 1200
\font\twelvebsy=cmbsy10 scaled 1200

\font\elevenrm=cmr10 scaled 1095
\font\eleveni=cmmi10 scaled 1095
\font\elevensy=cmsy10 scaled 1095
\font\elevenbf=cmbx10 scaled 1095
\font\eleventt=cmtt10 scaled 1095
\font\elevenit=cmti10 scaled 1095
\font\elevensl=cmsl10 scaled 1095
\font\elevensc=cmcsc10 scaled 1095
\font\elevenss=cmss10 scaled 1095
\font\elevenssi=cmssi10 scaled 1095
\font\elevenbmit=cmmib10 scaled 1095
\font\elevenbsy=cmbsy10 scaled 1095

\font\tensc=cmcsc10
\font\tenss=cmss10
\font\tenssi=cmssi10
\font\tenbmit=cmmib10
\font\tenbsy=cmbsy10

\font\ninerm=cmr9    \font\eightrm=cmr8    \font\sixrm=cmr6
\font\ninei=cmmi9    \font\eighti=cmmi8    \font\sixi=cmmi6
\font\ninesy=cmsy9   \font\eightsy=cmsy8   \font\sixsy=cmsy6
\font\ninebf=cmbx9   \font\eightbf=cmbx8   \font\sixbf=cmbx6
\font\ninett=cmtt9   \font\eighttt=cmtt8
\font\nineit=cmti9   \font\eightit=cmti8   
\font\ninesl=cmsl9   \font\eightsl=cmsl8
\font\niness=cmss9   \font\eightss=cmss8
\font\ninessi=cmssi9 \font\eightssi=cmssi8


\skewchar\twelvei='177    \skewchar\eleveni='177     \skewchar\ninei='177
\skewchar\eighti='177     \skewchar\sixi='177
\skewchar\twelvesy='60    \skewchar\elevensy='60     \skewchar\ninesy='60
\skewchar\eightsy='60     \skewchar\sixi='60


\catcode`@=11
\newskip\ttglue
\parindent=3em

\def\twelvept{\def\rm{\fam0\twelverm}
   \textfont0=\twelverm \scriptfont0=\ninerm \scriptscriptfont0=\sevenrm%
   \textfont1=\twelvei  \scriptfont1=\ninei  \scriptscriptfont1=\seveni%
   \textfont2=\twelvesy \scriptfont2=\ninesy \scriptscriptfont2=\sevensy%
   \textfont3=\tenex    \scriptfont3=\tenex  \scriptscriptfont3=\tenex%
   \textfont\itfam=\twelveit   \def\it{\fam\itfam\twelveit}%
   \textfont\slfam=\twelvesl   \def\sl{\fam\slfam\twelvesl}%
   \textfont\ttfam=\twelvett   \def\tt{\fam\ttfam\twelvett}%
   \textfont\bffam=\twelvebf   \scriptfont\bffam=\ninebf%
      \scriptscriptfont\bffam=\sevenbf   \def\bf{\fam\bffam\twelvebf}%
   \def\oldstyle{\fam1 \twelvei}%
   \tt \ttglue=.5em plus.25em minus.15em%
   \normalbaselineskip=13pt plus.5pt minus.5pt%
   \def\doublespacing{\baselineskip=26pt plus.5pt minus 1pt}%
   \def\spaceandahalf{\baselineskip=19.5pt plus.5pt minus .5pt}%
   \setbox\strutbox=\hbox{\vrule height9pt depth4pt width0pt}%
   \let\sc=\twelvesc   \let\big=\tenbig%
   \let\ss=\twelvess   \let\ssi=\twelvessi%
   \let\bmit=\twelvebmit\let\bsy=\twelvebsy%
   \normalbaselines\rm}

\def\elevenpt{\def\rm{\fam0\elevenrm}
   \textfont0=\elevenrm \scriptfont0=\eightrm \scriptscriptfont0=\sixrm%
   \textfont1=\eleveni  \scriptfont1=\eighti  \scriptscriptfont1=\sixi%
   \textfont2=\elevensy \scriptfont2=\eightsy \scriptscriptfont2=\sixsy%
   \textfont3=\tenex    \scriptfont3=\tenex  \scriptscriptfont3=\tenex%
   \textfont\itfam=\elevenit   \def\it{\fam\itfam\elevenit}%
   \textfont\slfam=\elevensl   \def\sl{\fam\slfam\elevensl}%
   \textfont\ttfam=\eleventt   \def\tt{\fam\ttfam\eleventt}%
   \textfont\bffam=\elevenbf   \scriptfont\bffam=\eightbf%
      \scriptscriptfont\bffam=\sixbf   \def\bf{\fam\bffam\elevenbf}%
   \def\oldstyle{\fam1 \eleveni}%
   \tt \ttglue=.5em plus.25em minus.15em%
   \normalbaselineskip=12pt plus.5pt minus.5pt%
   \def\doublespacing{\baselineskip=24pt plus.5pt minus1pt}%
   \def\spaceandahalf{\baselineskip=18pt plus.5pt minus .5pt}%
   \setbox\strutbox=\hbox{\vrule height8.5pt depth3.5pt width0pt}%
   \let\sc=\elevensc   \let\big=\tenbig
   \let\ss=\elevenss   \let\ssi=\elevenssi%
   \let\bmit=\elevenbmit\let\bsy=\elevenbsy%
   \normalbaselines\rm}

\def\tenpt{\def\rm{\fam0\tenrm}
   \textfont0=\tenrm \scriptfont0=\sevenrm \scriptscriptfont0=\fiverm%
   \textfont1=\teni  \scriptfont1=\seveni  \scriptscriptfont1=\fivei%
   \textfont2=\tensy \scriptfont2=\sevensy \scriptscriptfont2=\fivesy%
   \textfont3=\tenex \scriptfont3=\tenex   \scriptscriptfont3=\tenex%
   \textfont\itfam=\tenit   \def\it{\fam\itfam\tenit}%
   \textfont\slfam=\tensl   \def\sl{\fam\slfam\tensl}%
   \textfont\ttfam=\tentt   \def\tt{\fam\ttfam\tentt}%
   \textfont\bffam=\tenbf   \scriptfont\bffam=\sevenbf%
      \scriptscriptfont\bffam=\fivebf   \def\bf{\fam\bffam\tenbf}%
   \def\oldstyle{\fam1 \teni}%
   \tt \ttglue=.5em plus.25em minus.15em%
   \normalbaselineskip=11pt plus.5pt minus.5pt%
   \def\doublespacing{\baselineskip=22pt plus.5pt minus 1pt}%
   \def\spaceandahalf{\baselineskip=16.5pt plus.5pt minus .5pt}%
   \setbox\strutbox=\hbox{\vrule height8.5pt depth3.5pt width0pt}%
   \let\sc=\tensc   \let\big=\tenbig%
   \let\ss=\tenss   \let\ssi=\tenssi%
   \let\bmit=\tenbmit\let\bsy=\tenbsy%
   \normalbaselines\rm}

\def\ninept{\def\rm{\fam0\ninerm}
   \textfont0=\ninerm \scriptfont0=\sixrm \scriptscriptfont0=\fiverm%
   \textfont1=\ninei  \scriptfont1=\sixi  \scriptscriptfont1=\fivei%
   \textfont2=\ninesy \scriptfont2=\sixsy \scriptscriptfont2=\fivesy%
   \textfont3=\tenex  \scriptfont3=\tenex \scriptscriptfont3=\tenex%
   \textfont\itfam=\nineit   \def\it{\fam\itfam\nineit}%
   \textfont\slfam=\ninesl   \def\sl{\fam\slfam\ninesl}%
   \textfont\ttfam=\ninett   \def\tt{\fam\ttfam\ninett}%
   \textfont\bffam=\ninebf   \scriptfont\bffam=\sixbf%
      \scriptscriptfont\bffam=\fivebf   \def\bf{\fam\bffam\ninebf}%
   \def\oldstyle{\fam1 \ninei}%
   \tt \ttglue=.5em plus.25em minus.15em%
   \normalbaselineskip=10pt plus.5pt minus.5pt%
   \def\doublespacing{\baselineskip=20pt plus.5pt minus1pt}%
   \def\spaceandahalf{\baselineskip=15pt plus.5pt minus .5pt}%
   \setbox\strutbox=\hbox{\vrule height8pt depth3pt width0pt}%
   \let\sc=\sevenrm   \let\big=\ninebig%
   \let\ss=\niness    \let\ssi=\ninessi%
   \normalbaselines\rm}

\def\eightpt{\def\rm{\fam0\eightrm}
   \textfont0=\eightrm \scriptfont0=\sixrm \scriptscriptfont0=\fiverm%
   \textfont1=\eighti  \scriptfont1=\sixi  \scriptscriptfont1=\fivei%
   \textfont2=\eightsy \scriptfont2=\sixsy \scriptscriptfont2=\fivesy%
   \textfont3=\tenex   \scriptfont3=\tenex \scriptscriptfont3=\tenex%
   \textfont\itfam=\eightit   \def\it{\fam\itfam\eightit}%
   \textfont\slfam=\eightsl   \def\sl{\fam\slfam\eightsl}%
   \textfont\ttfam=\eighttt   \def\tt{\fam\ttfam\eighttt}%
   \textfont\bffam=\eightbf   \scriptfont\bffam=\sixbf%
      \scriptscriptfont\bffam=\fivebf   \def\bf{\fam\bffam\eightbf}%
   \def\oldstyle{\fam1 \eighti}%
   \tt \ttglue=.5em plus.25em minus.15em%
   \normalbaselineskip=9pt plus.5pt minus.5pt%
   \def\doublespacing{\baselineskip=18pt plus.5pt minus1pt}%
   \def\spaceandahalf{\baselineskip=13.5pt plus.5pt minus .5pt}%
   \setbox\strutbox=\hbox{\vrule height7pt depth2pt width0pt}%
   \let\sc=\sixrm   \let\big=\eightbig%
   \let\ss=\eightss \let\ssi=\eightssi%
   \normalbaselines\rm}

\def\tenbig#1{{\hbox{$\left#1\vbox to8.5pt{}\right.\n@space$}}}
\def\ninebig#1{{\hbox{$\textfont0=\tenrm\textfont2=tensy
   \left#1\vbox to7.25pt{}\right.\n@space$}}}
\def\eightbig#1{{\hbox{$\textfont0=\ninerm\textfont2=ninesy
   \left#1\vbox to6.5pt{}\right.\n@space$}}}

\let\singlespacing=\normalbaselines



\def\today{\ifcase\month\or
   January\or February\or March\or April\or May\or June\or
   July\or August\or September\or October\or November\or December\fi
   \space\number\day, \number\year}

\def\sciday{\number\day
   \space\ifcase\month\or
   January\or February\or March\or April\or May\or June\or
   July\or August\or September\or October\or November\or December\fi
   \space\number\year}

\newcount\tyme
\newcount\hour
\newcount\minute

\def\amorpm{a.m.}
\def\tod{\gettime\number\hour:\ifnum\minute<10{}0\fi\number\minute\space\amorpm}

\def\gettime{\tyme=\time
   \divide \tyme by 60
   \hour=\tyme
   \ifnum\hour=12\gdef\amorpm{p.m.}\fi
   \ifnum\hour=0 \advance \hour by  12\fi
   \ifnum\hour>12\advance \hour by -12\gdef\amorpm{p.m.}\fi
   \multiply \tyme by 60
   \advance \time by -\tyme
   \minute=\time
   \advance \time by \tyme}



\def\underule#1{$\setbox0=\hbox{#1} \dp0=\dp\strutbox
    \m@th \underline{\box0}$}


\def\narrow{\advance\leftskip by3em \advance\rightskip by3em}
\def\wide{\advance\leftskip by-3em \advance\rightskip by-3em}


\def\alph#1{\ifcase#1\or a\or b\or c\or d\or e\or f\or g\or h\or i\or j\or
   k\or l\or m\or n\or o\or p\or q\or r\or s\or t\or u\or v\or w\or x\or
   y\or z\fi}

\def\deg{\ifmmode^\circ\else$^\circ$\fi}


\def\applt{\mathrel{\mathpalette\@versim<}}
\def\appgt{\mathrel{\mathpalette\@versim>}}
\def\@versim#1#2{\lower2pt\vbox{\baselineskip0pt \lineskip-.5pt
   \ialign{$\m@th#1\hfil##\hfil$\crcr#2\crcr\sim\crcr}}}


{\catcode`p=12 \catcode`t=12 \gdef\\#1pt{#1}}
\let\getfactor=\\
\def\kslant#1{\kern\expandafter\getfactor\the\fontdimen1#1\ht0}
\def\vector#1{\ifmmode\setbox0=\hbox{$#1$}%
    \setbox1=\hbox{\the\scriptscriptfont1\char'52}%
	\dimen@=-\wd1\advance\dimen@ by\wd0\divide\dimen@ by2%
    \rlap{\kslant{\the\textfont1}\kern\dimen@\raise\ht0\box1}#1\fi}

\tenpt\singlespacing

\magnification=\magstep1
\vsize=9truein
\hsize=6.5truein
\hoffset=.25truein
\voffset=1truein

\centerline{\bf Optimizing the Zel'dovich Approximation}
\vskip .5in
\centerline{Adrian L. Melott, Todd F. Pellman, and Sergei
F. Shandarin}
\smallskip
\centerline{\it Department of Physics and Astronomy}
\centerline{\it University of Kansas}
\centerline{\it Lawrence, Kansas 66045 USA}
\vskip .5in
\baselineskip=12pt
\noindent {\bf ABSTRACT}
\medskip
We have recently learned that the Zeldovich approximation can be
successfully used for a far wider range of gravitational instability
scenarios than formerly proposed; we study here how to extend this
range.
In previous work (Coles, Melott and Shandarin 1993, hereafter CMS) we studied
the accuracy of
several analytic approximations to gravitational clustering in the mildly
nonlinear regime. We found that what we called the ``truncated Zel'dovich
approximation" (TZA) was better than any other (except in one case the ordinary
Zeldovich approximation) over a wide range from linear to mildly nonlinear
($\sigma \sim 3$) regimes. TZA was specified by setting Fourier amplitudes
equal to zero for {\it all} wavenumbers greater than $k_{n\ell}$, where
$k_{n\ell}$
marks the transition to the nonlinear regime.  Here, we study the
crosscorrelation of generalized TZA with a group of $n$--body
simulations for
three shapes of window function: sharp $k$--truncation (as in CMS), a tophat in
coordinate space, or a Gaussian. We also study the variation in the
crosscorrelation as a function of initial truncation scale within each type.
\medskip
We find that $k$--truncation, which was so much better than other things tried
in
CMS, is the {\it worst} of these three window shapes. We find that a Gaussian
window
$e^{-k^2/2k_G^2}$ applied to the intial Fourier amplitudes is the best choice.
It produces a greatly improved crosscorrelation in all cases we studied.
The optimum choice of $k_G$ for the Gaussian window is (a somewhat
spectrum--dependent) 1 to 1.5
times $k_{n\ell}$, where $k_{n\ell}$ is defined by (3).
Although all three windows produce similar power spectra and density
distribution functions after application of the Zeldovich approximation, the
agreement of the phases of the Fourier components with the $n$--body simulation
is
better for the Gaussian window. We therefore ascribe the success of the
best--choice Gaussian
window to its superior treatment of phases in the nonlinear regime. We also
report
on the accuracy of particle positions and velocities produced by TZA.
\vfill\eject
\noindent {\bf 1 INTRODUCTION}
\medskip
For nearly fifty years there has been interest in understanding the
gravitational growth of
density perturbations in an expanding universe. For the latter half of this
time, we have seen increasingly sophisticated numerical simulations performed
on increasingly powerful computers in an attempt to model this process. There
has
been a fruitful interaction with theory; much of the effort has gone into two
directions: (a) does a particular scenario produce something that looks like
our
universe? (b) what approximations can we develop to describe the general
properties of the clustering process? This paper lies in second of these
traditions.
\medskip
The Zel'dovich (1970) approximation is the focus of this paper. One of us (ALM)
suggests there are strong indications near the end of section 3 that he thought
this approximation might work for entropic perturbtions (hierarchical
clustering). However, it quickly was decided in a later paper (Zel'dovich 1973)
that it would
work only to
describe Universes in which large wavelength perturbations dominate, which were
associated with what was then called the ``adiabatic" picture or sometimes the
``pancake" model, after the large flattened structures that appeared in it.
\medskip
During the 80's, evidence gradually accumulated that the approximation had
wider validity. Filamentary structure appeared in a variety of numerical
simulations, beginning with CDM when Melott {\it et al.} (1983) found that it
had enhanced percolation.
\medskip
Coles, Melott and Shandarin (1993) hereafter CMS, conducted a series of tests
by
crosscorrelating $n$--body simulations with various approximate solutions to
the
same initial conditions. They found the Zel'dovich approximation, particularly
in a ``truncated" form implemented by smoothing the initial conditions to
remove
unwanted nonlinearity, was the most successful. The idea of the truncation of
the initial spectrum evolved from the very well known linear theory to the
comparison of $N$--body simulations having the same longwave perturbations but
different cutoffs as in Beacom {\it et al.} (1991) and Melott and Shandarin
(1993) then to the adhesion approximation as in Kofman {\it et al.} (1992), and
of the truncated Zeldovich approximation (CMS). In this paper we
improve on that success by finding the best way to do the initial smoothing. We
will
see that a considerable further improvement is made.
\medskip
 We
first define a dimensionless density contrast $\delta ({\bf x})$ for the matter
density $\rho$ in co--moving coordinates ${\bf x}={\bf r}/a(t)$ by
$$\delta({\bf x,t})={\rho({\bf x,t})-\rho_0\over \rho_0}\eqno (1)$$
$a(t)$ is the cosmological scale factor and, assuming a flat Universe
with $\Omega_0=1$ and in the absence of radiation and pressure terms,
$a(t)\propto t^{2/3}$ and $\rho_0(t)\propto a^{-3}\propto t^{-2}$. The
evolution of $\delta(\vec{\bf x},t)$ is described by the standard set of
equations (e.g. Peebles, 1980 Eq. (7.9) and (9.1)).
\medskip
Furthermore, we specify the following initial conditions. If $\delta_k$ is the
spatial Fourier transform of the density contrast (1), then our scale--free
initial perturbations are expressed by a power spectrum of the form
$$P(k)=<\mid\delta_k\mid^2>\propto k^n\; .\eqno(2)$$ In the following
discussion we will take $n=+1,0,-1,-2$ for illustrative examples. At the
initial time when the density contrast is everywhere small we assume that
phases of the Fourier components are randomly distributed on the interval
[0,2$\pi$]. In this case, $\delta({\bf x})$ is a Gaussian random field and all
of its statistical properties are completely contained in the power spectrum
(2).
\medskip
In Fourier space, components $\delta_k$ for large magnitudes $k$ correspond to
structure on small scales and similarly for small $k$ and large structures. We
define $k$--nonlinear, or $k_{n\ell}$, by
$$a^2(t)\int^{k_{n\ell}}_0P(k)d^3{\bf k}=1\; .\eqno (3)$$

With this definition of $k_{n\ell}$ we can say that for $k<k_{n\ell}$ we are
considering structures in the linear regime, that is, structures whose density
contrasts have grown approximately proportional to $a(t)$. Clearly, $k_{n\ell}$
decreases with time as larger scales become more nonlinear.
\bigskip
\noindent {\bf 2 A GENERALIZED TRUNCATED ZELDOVICH APPROXIMATION}
\medskip
With the Zeldovich approximation (Zeldovich 1970) we simply assign to each
material particle
(strictly to the particle's initial unperturbed Lagrangian co--ordinate ${\bf
q}$) a vector and move the particle along that vector. The Eulerian
(co--moving)
coordinate ${\bf x}$ of a particle at time $t$ is given by
$${\bf x}({\bf q},t)={\bf q}+a(t){\bf \nabla}\Phi_{i}({\bf q})\eqno (4)$$ where
$\Phi_i({\bf q})$ is the initial {\it velocity potential}, related to the
initial
gravitational potential by
$$\phi_{i}({\bf x})=-{3\over 2}H^2a^3\Phi_{i}({\bf x})\eqno (5)$$ so that
$\nabla^2\Phi_i\propto\delta$ and the approximation is readily obtained from
the initial conditions (2).
\medskip
The {\it ansatz} (4) leads to a catastrophe where trajectories cross, a
phenomenon known as shell crossing (e.g., Shandarin \& Zeldovich 1989).
However,
until the catastrophe is reached, the approximation performs well, but only for
$n\leq -3$, i.e., for spectra in which most of the power is concentrated
on large scales. This effect can be easily understood when one considers that
it
is on small scales that the highly nonlinear effects occur. Thus, we can
expect to improve upon the approximation (4) by first damping out initial power
for large $k$, i.e., small scales. We call this the ``truncated
Zel'dovich" approximation, or TZA.
\medskip
We investigate here the effect of three ``windows" applied in Fourier space.
That is, for a window $W(k)$ the initial conditions $\delta^*_k$
for the improved
approximation are just $\delta^*_k=W(k)\delta_k$ where $\delta_k$
are the initial conditions as given before (2). Since the phases of the
coefficients are not changed, we will be able to test directly the agreement.
\medskip
The window which we will refer to as $k$--truncated is simply
$$W_{tr}(k;k_{tr})=\cases{1,&$k\leq k_{tr}$\cr 0,&$k> k_{tr}$\cr}\eqno (6)$$
This has already been shown to be an improvement to the original approximation
(CMS), however, it was only investigated for
$k_{tr}=k_{n\ell}$. Here we tested it for a range from $k_{tr}=2k_f$ to
$k_{tr}=20k_f$ where $k_f$ is the fundamental mode of the box at two stages:
$k_{n\ell}=8k_f$ and $k_{n\ell}=4k_f$.
\medskip
The other two windows tested are a Gaussian window
$$W_G(k;k_{_G})=e^{-k^2/2k_G^2}\eqno (7)$$ and a top--hat window in real space
which corresponds in Fourier space to
$$W_{th}(k;R_{th})=3\biggl({\sin R_{th}k\over (R_{th}k)^3}-{\cos R_{th}k\over
(R_{th} k)^2}\biggr)\eqno (8)$$ We similarly tested these windows over a range
of $k_{_G}$ and $R_{th}$ to find the parameters for the best performance. The
meaning of ``best" will be clarified in section 4.
\medskip
We investigated a fourth window defined by
$$W(k;k^*)=\cases{1,&$k\leq k^*$\cr e^{-k^2/2k^{*2}},&$k>k^*$\cr}\eqno (9)$$
motivated
by the need to suppress small scale power in a gradual fashion and the belief
that the Zeldovich approximation worked well for large scale power and so we
should
leave those amplitudes unaffected. However, this window performed only slightly
better than $k$--truncated so we do not consider it further.
\bigskip
\noindent {\bf 3 NUMERICAL SIMULATIONS}
\medskip
The model data to which we compare the TZA approximations is
provided by a set of $N$--body experiments that approximate the evolution of a
cosmological density field with a set of particles on a grid with periodic
boundary conditions. The details are discussed much more completely by CMS and
in Melott and Shandarin (1993) hereafter MS; the
essentials of such a simulation are the following.
\medskip
Each $N$--body simulation is evolved from a set of initial density fluctuations
with power spectra of the form (2) and random phases. At very low amplitude our
use of the Zeldovich approximation (4) for initial conditions generates not
only particle displacements but also velocities in accord with the growing mode
of gravitational instability. The initial low amplitude restriction was such
that no particle could be displaced more than 1/2 the cell width from
homogeneity. We studied spectra corresponding to $n=1,0,-1,-2$, all generated
from the same set of random phases, which explains the similar overall
structures of simulations with different spectra. The simulations were run for
various expansion factors $a(t)$; we consider here only those stages
corresponding to $k_{n\ell}=8k_f$ and $k_{n\ell}=4k_f$. These two scales
represent a
good compromise between resolution in terms of particles which drives one to
large scales and the effect of our periodic boundary conditions which
leads one to small scales (Kauffmann \& Melott 1992).
\medskip
For all cases, $N$--body and various versions of TZA, we evolved $128^3$
particles,
each on a $128^3$ co--moving mesh with periodic boundary conditions. For the
$N$--body results we used the enhanced PM (particle--mesh) method of Melott
(1986). This makes them resolution equivalent to simulations with $128^3$
particles on a $256^3$ grid in traditional PM codes: see also Park (1990, 1991)
and
Weinberg et al. (1993).
Grey scale
plots of thin slices through these densities corresponding to $k_{n\ell}=8$
are shown in Figures 1a, 2a, 3a, and 4a  for the $N$--body simulations.
\bigskip
\noindent {\bf 4 CROSS CORRELATIONS}
\medskip
As in CMS
we use here the usual cross--correlation
coefficient to compare each grid--point in the resulting TZA
approximations to the corresponding grid--point in the $N$--body simulation.
This coefficient is given by
$$S={<(\delta_1\delta_2>\over \sigma_1\sigma_2}\eqno (10)$$ where $\delta_1$
and
$\delta_2$,
represent the density contrasts in the
TZA and $N$--body distributions, respectively;
$\sigma_i\equiv<\delta_i^2>^{1/2}$; and
averages are over the entire distribution. Note $\mid S\mid\leq 1$ and that
$S=+1$ implies that $\delta_1=C\delta_2$, with $C$ constant for every pixel.
\medskip
We exploit our use of identical phases before $n$--body evolution or
application of
an approximation to demand good agreement. Our crosscorrelation test would be
impossible without it.
\medskip
CMS found that the cross correlations between TZA
and $N$--body for different realizations of the random phases
agreed to the order of
$10^{-3}$. This allows us to make general conclusions about the performance
of an approximation from our investigation of only one realization.
\medskip
The cross--correlation technique applied to the ``raw" density fields may be
too strict a test. If the relative position of the structure is
very similar to the $N$--body results but is slightly displaced
from those of the $N$--body, a small cross--correlation can result. We can
overcome this by smoothing the resulting density fields by a convolution with a
Gaussian:
$$\delta({\bf x}, R)={1\over \biggl(\sqrt{2\pi} R\biggr)^3}\int \delta ({\bf
x}')\exp\biggl({-\mid{\bf x}-{\bf x}'\mid^2\over 2R^2}\biggr)d^3{\bf x}'\;
.\eqno (11)$$
Thus, the question becomes not how well do two density fields correlate, but
how fast do the correlations converge to unity with smoothing?
\medskip
Fields evolved from different values of $n$ will respond differently to a given
smoothing length $R$, so we find it more convenient to express $S$ as a
function of the rms density contrast $\sigma$. This is somewhat more intuitive;
$\sigma$ goes to zero as the field is more smoothed and $S$ goes to unity.
\medskip
We found when doing these crosscorrelations that there was no ambiguity in the
ordering of windows; the rank--ordering of the crosscorrelations did not change
with smoothing lengths for a given window size. We were thus able to explore a
wide variety of values for the three window scales $k_{tr}, k_G,$ and $k_{th}$.
(Although the tophat window is more naturally represented by a smoothing radius
$R_{th},$ for uniformity of notation we choose to use $k_{th}=1/R_{th}$ to
represent it.)
\bigskip
\noindent {\bf 5 RESULTS AND DISCUSSION}
\medskip
All window functions had a single, fairly broad maximum crosscorrelation for a
preferred value of $k$. We used the TZAs to generate a mass distribution for
this choice of best value for each window, and made plots of them in analogy
with the $n$--body plots. These are shown in Figures 2, 3, and 4. Each Figure
contains one $n$--body simulation and the three forms of TZA we compare with
it.
\medskip
All the pictures have a family resemblance, as expected. The arrangement of
the pictures in a square array allows the reader to easily compare appearance
across approximations.
As usual, more
negative $n$ leads to a more filamentary appearance in all cases and for all
indices the various versions TZA exaggerates the filamentarity. This is because
the TZA cannot follow in detail the highly nonlinear process of the breakup of
filaments into subclumps. The visual differences between the versions of TZA
are
more subtle. They resemble each other more for more negative $n$; this is
reasonable since they differ primarily in the treatment of larger wavenumbers
in the initial conditions and these are less important for more negative $n$.
Within a given index, the Gaussian window appears to produce a picture which
has a more ``smooth" or ``regular" appearance, whereas the others give an
impression of ``choppiness" or ``irregularity". Also in the Gaussian version
both the dark condensations and the grey filaments connecting them appear more
compact.
\medskip
In Figure 5 we
demonstrate the best--choice scale $k_w$ for each window in units of
$k_{n\ell}$.
The first thing we notice is that the optimum window value $k_w$ is nearly the
same for the two nonlinear stages (open and filled symbols in Figure 5),
reinforcing the fact that our results are not limited by resolution or boundary
conditions. Figures 2--4 show the stage $k_{n\ell}=8k_f$ which allows us to see
more structure within one slice.
\medskip
Figure 5 shows the range of best $k_w$ for our three window functions. The
small disagreement seen between the two stages is an artifact of the fact that
we only checked certain discrete values of $k_w$. The fact that the best value
of $k_w$ varies only a little for a given window over the range of indices (the
greatest being a
factor of about 5/3 for $k$--truncation), combined with the fact that we found
that $\pm$20\% error in $k_w$ makes very little difference, suggests that it
will be possible to state a general prescription good for all kinds of spectra.
\medskip
Figure 6 shows the crosscorrelations between the $n$--body simulations and our
best choice $k_w$ approximations for various spectra at both stages. Again all
the dependence on
particulars of the simulation box are removed by plotting against $\sigma$ of
the smoothed simulation. Each point is generated by crosscorrelating the
$n$--body
and the approximation both smoothed by convolution with the same Gaussian
windows. This smoothing of our results should not be confused with the window
function applied to the initial conditions.
\medskip
We can now compare the best results for each window against one
another. (The reader may also refer to Figure 6 in CMS.) The reader should note
that the crosscorrelation has no absolute meaning; the raw number depends on
the pixellization size relative to the size of structures after smoothing, and
this is different at the two stages. But the relative order should stay the
same, and does. A value of 1 would mean perfect agreement at the resolution of
one pixel, even if mass were rearranged inside pixels. CMS used the same
pixels,
so that can be directly compared to results here for the same $k_{n\ell}$ an
spectral index.
\medskip
For all eight panels in Figure 6, the Gaussian window produces the strongest
crosscorrelation between the resulting TZA density and that from the $n$--body
simulation. The $k$--truncation method, which was so successful compared with
other things tried by CMS, is the worst here. The greatest amount of
improvement found by CMS for TZA over straightforward use of the Zeldovich
approximation ($W=1$) was found for more positive indices. We  find here
that the transition to a Gaussian window also makes the greatest increment in
$S$ beyond the CMS result for more positive indices.
For $n=1$, $\sigma=1$, the value found by CMS was about 0.65; here it is about
0.85.
\medskip
Returning to Figure 5 momentarily, we observe that the best value of
$k_G/k_{n\ell}$ varies rather slowly with spectral index, lying in the range 1
to 1.5. One may be concerned about how to apply this to non--power law spectra
such as Cold Dark Matter. We speculate that the local slope at $k_{n\ell}$ will
determine this, and plan a check in the future. However, given the fact that
$\pm20$\% in $k_G$ makes little difference, we can recommend  generic use of
$k_w\sim 1.25k_{n\ell}$ for non--power law spectra in which all of the
quasilinear
regime of the spectrum lies in the range $-2\leq n\leq +1$. This includes
nearly all models of cosmological interest at this time. Modes $k<<k_{n\ell}$
will
be unaffected by our window, and modes $k>3k_{n\ell}$ will be damped to
insignificance.
\medskip
We can speculate based on the pictures why the Gaussian window works best. The
first crisis and failure of any Zeldovich approximation--based scheme happens
when trajectories cross. In real nonlinear gravity, the particles are slowed by
the attraction of the stream they have passed, which is ignored in the
approximation. This forces us to take out highly nonlinear modes. On the other
hand, they can help to preserve detail. It appears that the Gaussian window
works to balance these, reducing the amplitude of the more nonlinear modes
gradually as they begin to lead mass elements further astray. This can account
for the more focused and crisp appearance of the Gaussian based pictures.
\medskip
The Zeldovich approximation does not conserve either the power spectrum or
phases of Fourier components; it definitely includes some nonlinear mode
coupling. It is therefore useful to examine the agreement with the $n$--body
simulation. The power spectrum (or autocorrelation) is the most widely used
statistic in large--scale structure. We examine that first without additional
smoothing after applying TZA, in Figure 7. In all
cases, the nonlinear power is too low in all approximate schemes. The other
obvious point is that the spectra of the nonlinear approximations for a given
model are all
very similar in spite of the fact that their progenitors had different
spectra. This is a reflection of the fact that the nonlinear transfer of power
from small to large wavenumbers is dominant, as has been observed before. More
importantly for our purposes here, it shows us that the better agreement of the
Gaussian TZA cannot be a result of a spectrum closer to the $n$--body result.
In
fact, for $n=-1$ initial conditions, its spectrum is one of the farthest from
the $n$--body
result.
\medskip
We therefore look to phase differences. Each Fourier coefficient in the sum
that describes our density field has an amplitude and a phase angle
$\alpha:\delta_{\bf k}=\mid\delta_{\bf k}\mid e^{i\alpha}$. We can measure the
angle
$\theta=\alpha_N-\alpha_Z$ between the phases in the $n$--body simulation and
the
approximation to it. Perfect agreement would imply cos $\theta=1$;
anticorrelation of phases cos $\theta=-1$; and for randomized phases cos
$\theta$
would average 0. We expect the phase agreement to deteriorate with increasing
$k$; we thus average cos $\theta$ within spherical shells of $k$ and plot
$<\cos\theta>$ as a function of $k$ in Figure 8. The results are in agreement
with our crosscorrelation study: the Gaussian based  approximation has phases
which agree the best with those of the $n$--body simulation, and this agreement
is weakly if at all spectrum dependent. The $k$--truncation based approximation
is the worst and the most spectrum dependent. This is in perfect accordance
with the fact that we have improved TZA and greatly reduced its spectral
dependence by using a Gaussian window, and reinforces the importance of phase
information as stressed by Scherrer et al. (1991), Ryden and Gramman (1991),
and
Howe (1993).
\medskip
We can understand the performance a little better by examining the density
distribution function. In Figure 9 we show the number of cells  $N$ found with
mass density $\rho$ (in units of the mean). In all cases the approximations
underestimate the number of high density cells and overestimate the number of
lower density cells. We can also see that there is no systematic difference
between the windows. Therefore the difference in crosscorelation amplitude must
depend primarily on producing the correct location of mass condensations,
rather
than substantial differences in their density contrast.
\medskip
It is worth commenting that our results do not imply that Gaussian smoothing is
the best for restoring initial conditions from our nonlinear universe with the
Zeldovich approximations; smoothing does not commute with the approximation.
Melott (1993) has shown that if one wishes to smooth an evolved state in
preparation for computing its linear precursor, then $k$--truncation works
best.
This is probably because the sharp truncation effectively removes nonlinearly
generated modes which are of higher order than the Zeldovich approximation and
would thus create a false signal when mapping back to the initial conditions;
the signal would be false regardless of their amplitude. When extrapolating
forward, as we are studying here, the effect of ``sticking" in pancakes can
apparently be mimicked by a gradual reduction of amplitude with increasing $k$.
\medskip
Although we have conducted a fairly broad search, there are an infinite number
of possible smoothing windows and we cannot exclude the possibility that some
untried one would be even better than Gaussian. But it seems that finding it
would be difficult if not impossible without a specific prediction based on
theory.
\medskip
A substantial improvement now exists as compared with linear theory, as one can
see by comparing the crosscorrelation amplitudes we get from Gaussian TZA with
those derived from linear theory. For the most challenging $n=+1$ spectra, we
improve the correlation from 0.6 to 0.85 at $\sigma_\rho=1$ and from about 0.4
to better than 0.75 at $\sigma_\rho=2$. For $n=-1$, close to the slope on
scales
going nonlinear today, we see an improvement from 0.85 to about 0.95 for
$\sigma_\rho=1$, and from 0.75 to 0.85 for $\sigma_\rho=2$. We have removed
much
of the spectrum dependence found in the CMS version of TZA, and it is now much
better than linear theory for all spectral indices.
\medskip
Much of the analytic theory of large--scale structure is based on the idea of
smoothing to linearity, then using linear perturbation theory or simple
extensions of it. Our results show that any calculations which can be based on
TZA will be in much closer agreement with reality.
\medskip
After this paper was submitted, we completed similar analyses of the
frozen--flow
approximation (Melott {\it et al.} 1994a) and the adhesion approximation
(Melott {\it et al}, 1994b). Although these are considerably more complicated,
they both crosscorrelated substantially worse than TZA. The adhesion
approximation was better for some statistical quantities such as the mass
density distribution function and the power spectrum, but worse dynamically in
the sense of moving mass to the right place.
\medskip
Second--order Lagrangian perturbation theory (ZA may be considered as first
order) has recently been found by Melott {\it et al.} (1994c) to constitute a
slight improvement over TZA, {\it if} the initial conditions are truncated by
an optimal Gaussian smoothing. The improvements over first--order TZA are
rather small; it is
a question of taste whether it is worth the moderate complication.
\medskip
We have not yet completed a similar analysis of the linear evolution of
potential approximation (Brainerd {\it et al.} 1994; Bagla and Padmanaban
1994). This might do rather well. However, we
wish to point out that this is not really an analytic nonlinear approximation,
but rather a different way of doing $N$--body simulations. It consists of
moving
particles around over timesteps while assuming that the background potential is
content, {\it i.e.}, evolves according to linear perturbation theory. In
practice this is almost as expensive as doing a full $N$--body simulation, and
it
cannot be done analytically. It is therefore not directly comparable with TZA,
which can be written analytically and executed in what is equivalent to one
timestep of an $N$--body simulation.
\medskip
Since TZA works so well, at the request of a referee we have also examined
the distribution of errors in particle positions and velocities as compared
with
$N$--body. We define the position error
$$\Delta  x={\mid\bar {\bf x}_{TZA}-\bar {\bf x}_{N-b}\mid\over
\lambda_{n\ell}}\eqno (12)$$ where
$\lambda_{n\ell}$ is the nonlinearity wavelength. Figure 10 shows a histogram
of $\Delta x$. A typical position error is spectrum dependent: $\Delta x\sim
0.15\lambda_{n\ell}$ for $n=+1$ and $\Delta x\sim 0.075 \lambda_{n\ell}$ for
$n=-2$, which is in a good qualitative agreement with all previous results.
\medskip
The velocity field is a resolution -- dependent quantity, and cannot be
reported independent of some assumed smoothing window. In most practical
applications, approximations like TZA are used in the quasi--linear regime,
between the domain of validity of Eulerian perturbation theory and the fully
nonlinear regime best handled by $N$--body simulations. We therefore choose to
bin
the velocities to define a center--of--mass velocity for our 128$^3$
density pixels. This density field is then smoothed by a Gaussian (11) for
which the resulting RMS density contrast is unity. This is an extremely stable
measure (about $R=4h^{-1}$ Mpc for galaxy data). We report
$$\Delta v={\mid\bar {\bf v}_{TZA}-\bar {\bf v}_{N-b}\mid\over
H\lambda_{n\ell}}\eqno (13)$$ where
$H$ is the Hubble expansion parameter at the moment under analysis. In Figure
11 we show the distribution of $\Delta v$, weighted by mass. The dependence on
spectrum is much weaker than in the position error. Since both position and
velocity errors are given in dimensionless (nautral) units they can be compared
with each other. The velocity errors are considerably smaller, which probably
can
be related to the smoothing of the velocity distribution. In passing we note
that
the Zeldovich approximation itself is more accurate in terms of velocities than
coordinates (Doroshkevich, Ryabenkii, and Shandarin (1973).

\bigskip
\noindent {\bf 7 CONCLUSIONS}
\medskip
We have conducted a controlled study of the truncated Zeldovich approximation,
which CMS found worked in a spectrum--dependent fashion but always better than
linear theory. The TZA approximation consists of multiplying the linear Fourier
coefficients by a window function $W(k/Ck_{n\ell})$ where $C$ is a constant to
be
determined and $k_{n\ell}$ marks the transition to the nonlinear regime. We
explored three shapes for the function $W$: a step function, a Gaussian, and
the Fourier transform of a tophat (uniform sphere) and we varied $C$ for each
$W$.
\medskip
 We found
that:
\smallskip
(a) A Gaussian window $e^{-k^2/2k_G^2}$ produces the best crosscorrelation with
$n$--body simulations.
\smallskip
(b) The best choice for $C$ for a Gaussian window is in the range 1 to 1.5,
depending on the
spectral index of the initial conditions, but very little error will be
introduced by using 1.25 for all cases in the range $-2\leq n\leq1$.
\smallskip
(c) Using this window dramatically improves the performance for the more
challenging positive--$n$ case, removing much of the spectral dependence found
in
CMS.
\smallskip
(d) The reason for better performance of the Gaussian window is based on more
nearly correct phases of Fourier coefficients in the nonlinear regime, whereas
the power spectrum and the density distribution function
produced are nearly the same for all windows. Visually all windows produce
quite similar distributions.
\smallskip
(e) The use of TZA still  considerably
underestimates the power at large $k (k>k_{n\ell})$ and the density
counts at high densities ($\rho\appgt 6$).

\bigskip
\noindent {\bf  ACKNOWLEDGEMENTS}
\medskip
We are grateful for the financial support of (USA) NSF grants AST--9021414,
OSR--9255223, and NASA grant NAGW--2923, which made this work possible. One of
us (TP) is especially grateful from NSF Research Experiences for
Undergraduates supplemental grant. Our large--scale computations were performed
on a Cray--2 and a Convex C3 at the National Center for Supercomputing
Applications, Urbana, Illinois, USA.
\bigskip
\noindent {\bf REFERENCES}
\def\ref{\par\noindent\hangindent\parindent\hangafter1}
\bigskip
\ref
Bagla, J.S., and Padmanabhan, T., 1994 MNRAS, in press.
\medskip
\ref
Beacom, J., Dominik, K., Melott, A., Perkins, S., and Shandarin, S., 1991, ApJ
372, 351
\medskip
\ref
Brainerd, T.G., Scherrer, R.J., and Villumsen, J.V., 1994, ApJ in press
\medskip
\ref
Coles, P., Melott, A.L., and Shandarin, S.F., 1993, MNRAS 260, 765
\medskip
\ref
Doroshkevich, A.G., Ryabenkii, V.A., and Shandarin, S.F., 1973, Astrophysics 9,
144
\medskip
\ref
Howe, N.R., 1993, PhD Thesis, Princeton
\medskip
\ref
Kofman, L., Pogosyan,  S., Shandarin, S.F., and Melott, A.L., 1992, ApJ 393,
437
\medskip
\ref
Kauffmann, G., and Melott, A.L., 1992, ApJ 393, 415
\medskip
\ref
Melott, A.L., 1993, ApJ Lett, 414, L73
\medskip
\ref
Melott, A.L., 1986, Phys. Rev. Lett. 56, 1992
\medskip
\ref
Melott, A.L., Einasto, J., Saar, E., Suisalu, I., Klypin, A.A., and Shandarin,
S.F., 1983, Phys. Rev. Lett. 51, 935
\medskip
\ref
Melott, A.L., Lucchin, F., Matarrese, S., and Moscardini, L., 1994a, MNRAS
000,000
\medskip
\ref
Melott, A.L., and Shandarin, S.F., 1993, ApJ 410, 469
\medskip
\ref
Melott, A.L., Shandarin, S.F., and Weinberg, D.H., 1994b, ApJ in press
\medskip
\ref
Melott, A.L., Buchert, T., and Weiss, A., 1994c, Astron. Ap., in
preparation
\medskip
\ref
Melott, A.L., and Shandarin, S.F., 1993, ApJ 410, 469
\medskip
\ref
Park, C., 1990, PhD Thesis, Princeton
\medskip
\ref
Park, C., 1991, MNRAS 251, 167
\medskip
\ref
Ryden, B., and Grammann, M., 1991, ApJ 383, L33
\medskip
\ref
Scherrer, R.J., Melott, A.L., and Shandarin, S.F., 1991, ApJ 377, 29
\medskip
\ref
Shandarin, S.F., and Zeldovich, Ya.B., 1989, Rev.Mod.Phys. 61, 185
\medskip
\ref
Weinberg, D., et al., 1993, in peparation
\medskip
\ref
Zeldovich, Ya.B., 1970, A\&A 5, 84
\medskip
\ref
Zeldovich, Ya.B., 1973, Astrophysics 6, 164
\vfill\eject
\noindent {\bf FIGURE CAPTIONS}
\medskip
\ref
{\bf Figure 1.} A grayscale plot of thin ($L/128$) slices of the simulation
cube, and the approximations to it for index $n=+1$ initial conditions at stage
$k_{n\ell}=8$.
(a) the $n$--body simulation
(b) the k--truncated TZA approximation (c) the tophat--truncated TZA model (d)
the Gaussian--truncated TZA model.
\medskip
\ref
{\bf Figure 2.} As in Figure 1, but for $n=0$ initial conditions.
\medskip
\ref
{\bf Figure 3.} As in Figure 1,  but for $n=-1$ initial conditions.
\medskip
\ref
{\bf Figure 4.} As in Figure 1, but for $n=-2$ initial conditions.
\medskip
\ref
{\bf Figure 5.} A plot of the value of $k_w/k_{n\ell}$ which gave the best
crosscorrelation for each choice of window function. The solid figures are for
stage $k_{n\ell}=4k_f$ the open for stage $k_{n\ell}=8k_f$. The hexagons
represent the
value for the tophat window, the squares the gaussian window, and the triangles
the $k$--truncation window. Many open figures are apparently missing because
they coincide
with the same figure filled.
\medskip
\ref
{\bf Figure 6.} A plot of the crosscorrelation $S$ as defined in the text
between
the best TZA generated density field (Figs. 2--4) and the full $n$--body
simulations against the rms density fluctuation in the simulation. Both are
smoothed by the same Gaussian window. Solid line: Gaussian window.
Dot--dashed line: tophat window. Dashed line: $k$--truncation (a) for
$k_{n\ell}=8k_f$ (b) for $k_{n\ell}=4k_f$.
\medskip
\ref
{\bf Figure 7.} Power spectra for the various $n$--body simulations at
$k_{n\ell}=8k_f$ (heavy
solid line) and for the best TZA with the $k$--truncation
window (dashed line), tophat window (dot--dashed line) and Gaussian window
(solid line).
\medskip
\ref
{\bf Figure 8.} The average effective phase angle error for each of the three
windows,
as measured by $<\cos\theta>$ as described in the text, all at stage
$k_{n\ell}=8k_f$. Simulation with $n=+1$: short--long dash line. $n=0$: short
dash.
$n=-1$: long dash. $n=-2$: dot--short dash.
\medskip
\ref
{\bf Figure 9.} The mass density distribution in terms of the number of cells
$N$ with
density $\rho$ in units of the mean density, with CIC binning of $128^3$
particles on our $64^3$ mesh.
\medskip
\ref
{\bf Figure 10.} A histogram of the difference in position for identical
particles as evolved by TZA or by $n$--body, divided by $\lambda_{n\ell}$.
\medskip
\ref
{\bf Figure 11.} A histogram of the difference in velocity for identical
particles as evolved by TZA or by $n$--body, divided by $H(z)\lambda_{n\ell}$.
\bye